\documentclass[nato]{crckbked}

\usepackage{graphicx}
\usepackage{klucite}
\makeindex

\begin{document}
\numreferences

\begin{article}

\begin{opening}
\title{$c$\/-axis Intra-Layer Couplings in the 
CuO$_2$ Planes of High-T$_c$ Cuprates}
\author{J. \surname{R\"ohler}\email{abb12@uni-koeln.de}}
\institute{Universit\"at zu K\"oln\\ 
Z\"ulpicher Str. 77, D-50937 K\"oln, Germany}

\runningauthor{J. R\"ohler}

\begin{abstract}
We discuss the static deformations of the doped CuO$_2$ lattices in 
YBa$_2$Cu$_3$O$_x$.  Intrinsic lattice effects driven by the strong 
correlations of the electron system are separated from extrinsic 
chemical and bandstructure effects. $c$\/-axis displacements 
of the planar copper atoms 
seem to be a generic property of the metallic CuO$_2$ lattices.

\end{abstract}
\end{opening}

\section{Introduction}
Unconventional metals exhibit a plethora of unusual lattice effects 
not observed in ``simple'' band metals.  In the physics of strongly 
correlated electron systems these puzzling ``lattice instabilities'' 
are usually ignored or relegated to the high energy scale of the 
crystal chemistry.  This lecture focusses on the $static$ distortions 
of the nearly square planar CuO$_2$ lattices in the under- and 
overdoped regimes of YBa$_2$Cu$_3$O$_x$, the superconducting cuprate 
with the best known lattice structure.  We shall discuss the doping 
dependence of the basal lattice parameters, the CuO$_2$ dimples, and 
the orthorhombicity.  The structural data suggest that doping into a 
Mott insulator leads to weak but significant deformations of the 
metallic CuO$_2$ lattice.  Our focus is on the structural aspects of 
the normal state; it is not intended to suggest the inclusion of 
electron-lattice interactions into a specific model of high-$T_c$ 
superconductivity.

\section{Planar and perpendicular orbitals}

The pioneering work of M\"uller and Bednorz \cite{BedMue86} has been 
guided by the idea of high-$T_c$ superconductivity (HTSC) in 
transition-metal chalcogenides with strong Jahn-Teller (JT) effects.  
In fact, the strong JT effect is confirmed to play a crucial role for 
the electronic structure of the superconducting cuprates selecting 
from the $e_{g}$ doublet the antibonding $3d_{x^2-y^2}$ as the $only$ 
relevant $d$ orbital in the plane of the Cu--O polyhedron.  It 
generates the ernormous stability of the square-planar CuO$_2$ 
structure.  An electronically active role of the perpendicular 
$3d_{z^2}$ orbital is however not confirmed.

But superconductivity is not a purely 2-dimensional effect.  $T_{c}$ 
depends strongly on interactions of the planes with the doping 
reservoir, and a mechanism linking the plane with its perpendicular 
surroundings is expected to be involved in the formation of the 
electronic ground state.  Recent calculations of the one-electron 
bandstructure of YBa$_2$Cu$_3$O$_7$ \cite{PavAnd} show that the 
generic low-energy layer-related features have to be described by the 
planar orbitals Cu$3d_{x^2-y^2}$, O$2p_{x,y}$ directing in the plane, 
$and$ the isotropic Cu$4s$ orbital.  The chemical trend of $T_{cmax}$ 
is found to be controlled by the energy of a ``perpendicular 
orbital'' , a hybrid between Cu$4s$, Cu$3d_{z^2}$, the apical O$2p_{z}$, 
and farther $d_{z^2}$ or $p_{z}$ orbitals.  Interestingly materials 
with the perpendicular orbital more localized onto the CuO$_2$ 
layers, $i.e.$ with pure Cu$4s$ character, exhibit the highest $T_{cmax}$. 
It seems that the strong hybridization between Cu$3d_{x^2-y^2}$ and Cu$4s$ 
allows for atomic displacements in the rigid 
covalent CuO$_2$ lattice of the parent Mott insulator such 
that doped holes are optimally accomodated in the background of 
the quantum liquid of spin singlets.    

Notably the 
dependence of $T_{cmax}$ from the structure and the chemistry 
perpendicular to the planes enters a single-band many electron 
Hamiltonian simply by the ratio of the $nnn$ and $nn$ hopping 
integrals, $t'/t$.  It is worthwile to stress that the (rare) 
acknowledgement of the chemistry outside the planes as an integral 
part of the HTSC problem does not justify dynamically independent 
treatments of the Cu$3d$ and O$2p$ electrons in any kind of multi-band 
many electron Hamiltonians.

\section{Dimpled CuO$_2$ planes}

The CuO$_2$ planes of the superconducting cuprates are not perfectly 
flat but dimpled.  Many diffraction work, $e.g.$ Ref.  
\cite{CavRup,SchKar,CavBee,FisRus}, and some EXAFS studies 
\cite{RoeLoe,RoeCru} yield evidence for dimples in the relevant 
CuO$_2$ planes of the multi-layer cuprates, $i.e.$ the planes adjacent 
to the doping reservoirs, $cf.$ table I. ``Buckled'' planes 
are also observed in single-layer materials, most clearly in the doped 
La$_2$CuO$_4$ systems, but so far not in the single-layer Hg- and Tl-cuprates \cite{SchKar}.  
As we shall see, the dimplings 
in the latter materials are expected to be $small$, and since their CuO$_2$ planes are 
crystallographic inflection planes the usual crystallographic refinements 
tend to average them to zero.

\begin{table}
\caption{CuO$_2$ dimplings in typical HTSC, see text.}
\begin{tabular}{lccccc} \hline
\label{tab1}
 Compound& $P_{\perp}$ & $N_{L}$ & Dimpling (\AA)& $T_{c}$& Ref.\\  \hline
YBa$_2$Cu$_3$O$_x$ & $3^+ - 2^+$& 2 & 0.28 & 92 & (3, 9)\\
YBa$_2$Cu$_4$O$_8$ & $3^+ - 2^+$& 2 & 0.24 & 80 & (8)\\
Pb$_2$Sr$_2$YCu$_3$O$_8$ & $3^+ - 2^+$& 2 & 0.23 & 70 & (6)\\
HgBa$_2$(Ca,Y)Cu$_2$O$_x$ & $2^+/3^+ - 2^+$& 2& 0.10 & 85-110 & (4)\\
HgBa$_2$CaCu$_2$O$_x$ & $2^+ - 2^+$& 2&0.02 & 120 & (4)\\
HgBa$_2$Ca$_{2}$Cu$_3$O$_x$ & $2^+ - 2^+$& 3&0.05 & 132 & (4)\\
HgBa$_2$Ca$_{3}$Cu$_4$O$_x$ & $2^+ - 2^+$& 4&0.05 & 127 & (4)\\
\hline
\end{tabular}
\end{table}

We distinguish ``chemical'' dimples driven by external electrostatic 
polarization along $c$, mixed valence of the dopants, lattice 
mismatches and other bandstructure effects, from 
``correlation-driven'' dimples generic of the strongly correlated 
electron system.  We show that the former define a $large$ lengthscale 
of $\sim 0.3$ {\AA}, and the latter a $small$ lengthscale of $\sim 
0.03$ {\AA}, corresponding to buckling angles of $\sim 10^{\circ}$, and 
$\sim 1^{\circ}$, respectively.  The contribution of the chemical 
dimples in many of the doped metallic phases can be traced back to the 
lattice of their antiferromagnetic parent phase, and thus be 
differentiated from the correlation-driven dimples.  However in 
concentrated alloys, doped by varying ratios of one or two 
combinations of heterovalent cations, the correlation-driven dimples 
may be masked by the varying external electrostatic polarizations \cite{ChmSha}.  
Table~I compares the CuO$_2$ dimplings in some typical 
multi-layer cuprates with each other: as a function of the static 
charge contrast polarizing the CuO$_2$ layer perpendicularly, 
$P_{\perp}$, the number of CuO$_2$ layers, $N_{L}$, and $T_c$.  The 
$large$ lengthscale of $\sim 0.3$ {\AA} shows clearly up in all 
compounds with $P_{\perp}=2^+-3^+$, but is absent in compounds with 
$P_{\perp}=2^+-2^+$.  Materials alloyed with heterovalent cations, 
$e.g.$ Ca$^{2+}$/ Y$^{3+}$, exhibit intermediate values.  Notably 
CuO$_2$ layers sandwiched between isovalent layers seem to exhibit 
higher $T_{cmax}$ than those between heterovalent layers, $cf.$ Ref.  
\cite{KamIsh}.

\begin{figure}[b]
\tabcapfont
\centerline{
\begin{tabular}{c@{\hspace{2.5cm}}c}
\includegraphics[width=5cm]{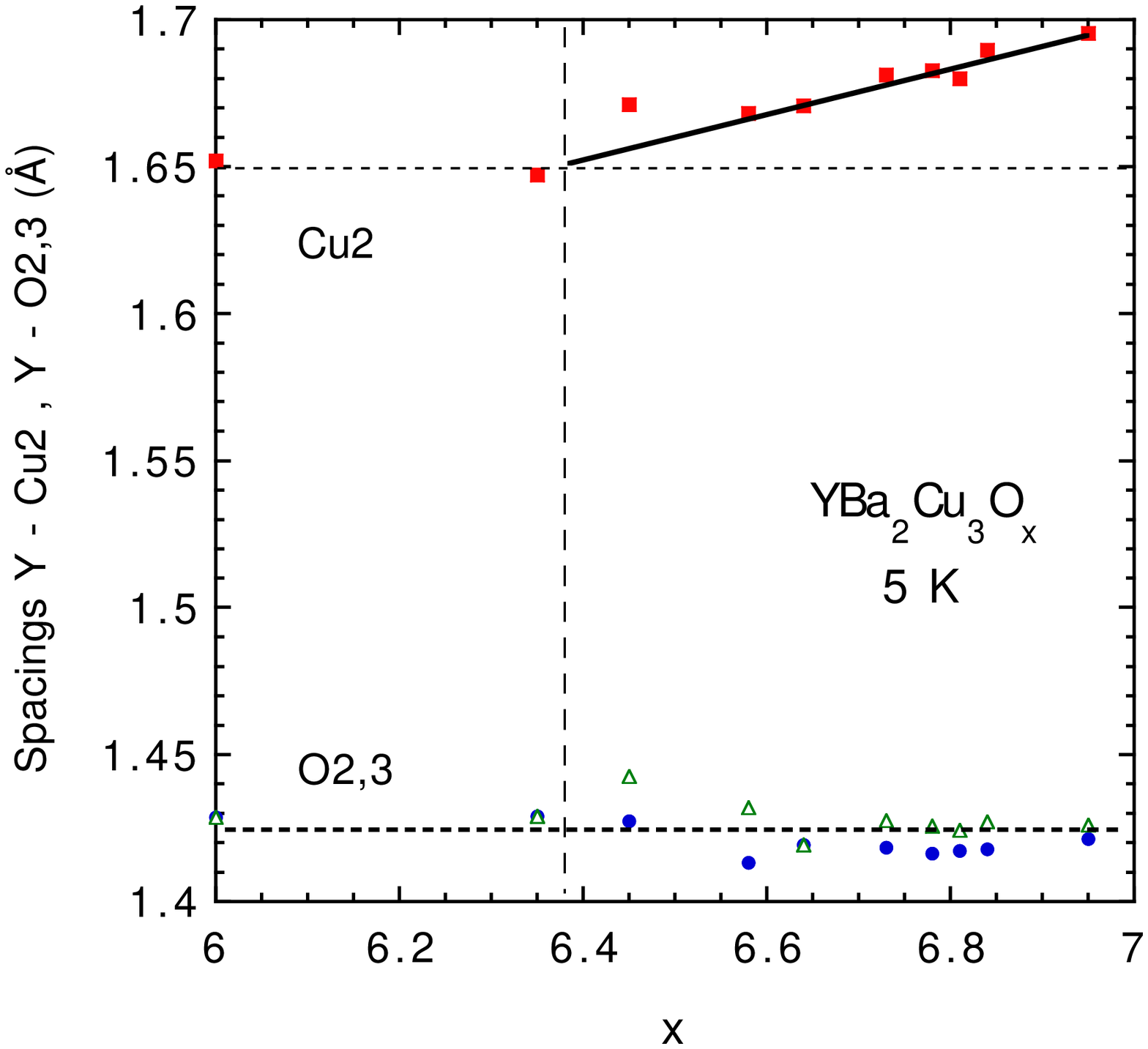} &
\includegraphics[width=3.cm]{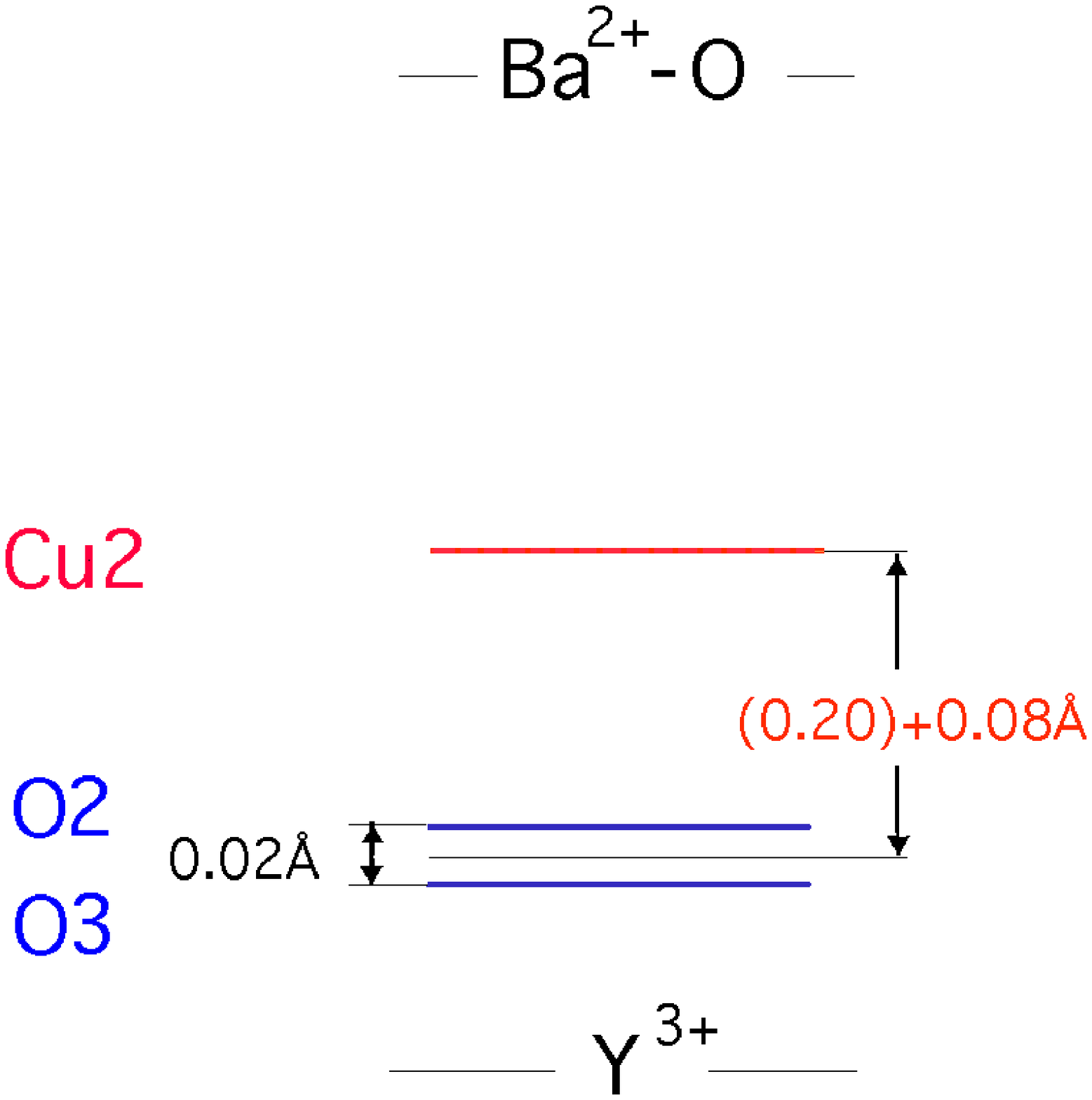} \\
a.~~ Y--Cu2, Y--O2,3 spacings & b.~~ CuO$_2$ intralayer spacings  
\end {tabular}}
\caption{Layer spacings in YBa$_2$Cu$_3$O$_x$.  a.: Y--Cu2, and Y--O2,3 
interlayer spacings as a function of oxygen concentration, $x$, at 5 K. After Cava $et$ 
 $al.$ (3). Open triangles: O2. Open circles: O3 . 
 b.: CuO$_2$ plane sketched as a stack of copper and 
 oxygen layers sandwiched between polarizing adjacent layers (arbitrary scale). }\label{spacings}
\end{figure}

\section{Perpendicular and planar CuO$_2$ deformations in YBa$_2$Cu$_3$O$_x$}

Figure~1b sketches a metallic CuO$_2$ layer of 
YBa$_2$Cu$_3$O$_x$ as a stack of copper and oxygen layers.  The 
horizontal dashed lines in Figure~1a indicate a 
``chemical'' offset of $\simeq 0.23$ {\AA}, typical for 
Ba$^{+2}$/Y$^{+3}$ polarization.  The spacing between the Cu2 and O2,3 
layers depends clearly on doping.  While the Cu2 layer starts to move 
towards the Ba--O layer by about 0.05 {\AA} between the onset of the 
metallic phase ($x\simeq 6.38$) and $x_{opt}$, the O2,3 layers seem to 
be almost unaffected.  We show in figure~3 that the scatter 
in the Y--O2, O3 data points does not behave arbitrarily, and that the 
relatively weak perpendicular displacements of O2 and O3 are correlated 
with the planar deformations.

\subsection{Lattice parameters and orthorhombic deformations}

The relatively strong orthorhombicity of its unit cell puts 
YBa$_2$Cu$_3$O$_x$ somewhat outside the lattice systematics of the 
most prominent high-$T_c$ materials, which have mostly tetragonal unit 
cells.  Weak orthorhombic strain might be however operative also on 
the nominally tetragonal lattices, if a correlation-driven anisotropy 
of the 2-D Fermi surface is a generic feature of HTSC, as 
theoretically suggested by renormalization group methods \cite{HalMet}.

\begin{figure}[b]
\tabcapfont
\centerline{
\begin{tabular}{c@{\hspace{1.5cm}}c}
\includegraphics[width=5cm]{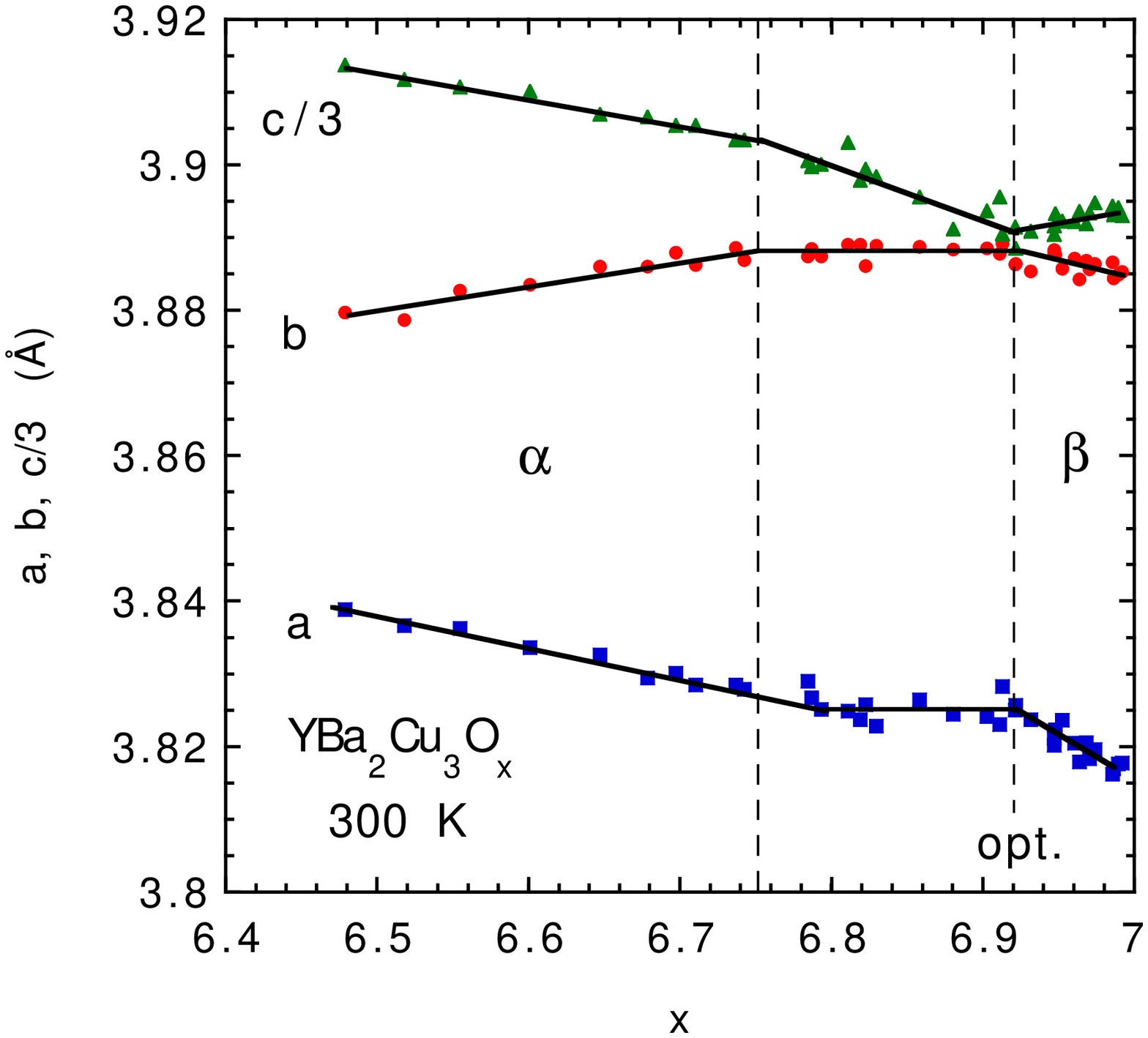} &
\includegraphics[width=5cm]{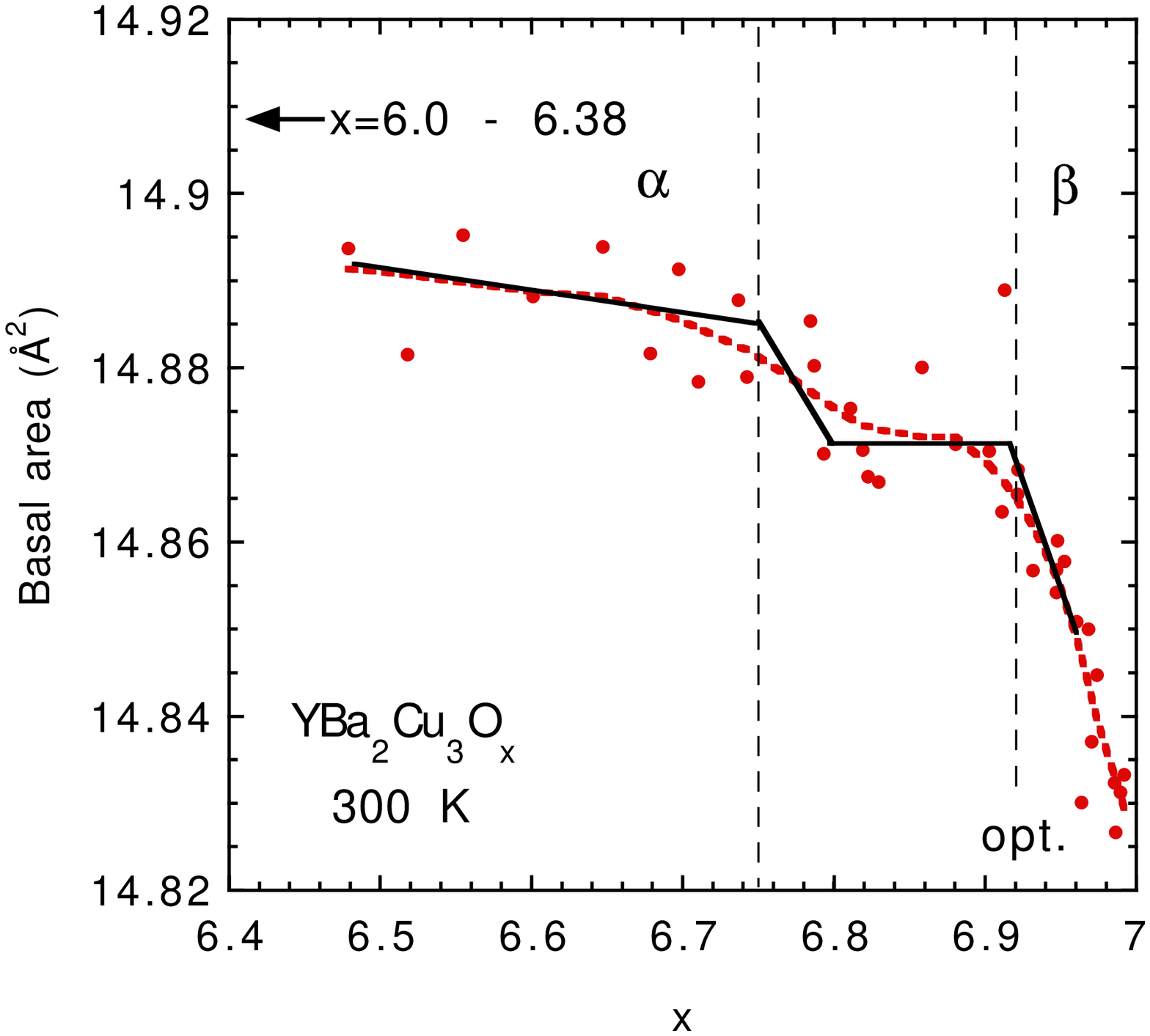} \\
a.~~ Lattice parameters & b.~~ Basal area
\end {tabular}}
\caption{a.: Lattice parameters $vs.$ oxygen concentration in 
near-equilibrium samples of metallic YBa$_2$Cu$_3$O$_x$ synthezised by 
direct oxydation of the elements (DO), after Conder $et$ $al.$ (5).  
Drawn out lines are guides to the eye.  $\alpha$ and 
$\beta$ label different orthorhombic strain ellipsis, see text.  b.: Basal area, 
$a\cdot b$, from the data in a.  Dashed line: smoothed average. The 
step indicates the onset of a possible two-phase regime.}\label{lattice}

\end{figure}
Figure~2a displays the doping dependence of the lattice 
parameters $a, b, {c/3}$ in the metallic phases of 
YBa$_2$Cu$_3$O$_x$.  $a\neq b$ in the under- and overdoped regimes, 
but the inequality results clearly from different types of 
orthorhombic strain ellipsis.  The quadrupolar $\alpha$\/-ortho 
stresses $a$ while straining $b$, and the monopolar $\beta$\/-ortho 
stresses both axis.  ``Plateaus'' of $a$ and $b$ between $x\simeq 6.8$ 
and 6.9 connect the two regimes suggesting a thermodynamical instable 
regime. If the orthorhombic strain originated exclusively from Cu1--O4 
hybridization in the 1-dimensional charge reservoir, we expect the 
$b$\/-, and the $a$\/-axis to exhibit similar orthorhombic 
deformations in both, the under- and overdoped regimes.  The abrupt 
change from the quadrupolar $\beta$\/-ortho to the monopolar 
$\beta$\/-ortho deformation at $x_{opt}$ points however to an 
additional and different mechanism to be operative.

%
Figure~2b displays the doping dependence of the basal 
area, $B(x)=a\cdot b$.  Evidently accomodation of oxygen atoms in the 
reservoir layer does not increase $B(x)$ as might be expected from 
simple stereochemical grounds, but reduces the basal area.  Hence the 
doping dependence of $B(x)$ is most likely controlled by the ground 
state of the quantum liquids in the metallic planes, and not by the 
anisotropic oxygen diffusion processes in the non-stoichiometric chain 
layer.  To discuss $B(x)$ in more detail we find it useful to 
associate $\kappa_{p}\propto -\partial B/\partial {n_{h}}$ with a 
2-dimensional ``electronic compressibility'', eliminating the almost 
doping independent bandstructure effects.  Here $n_{h}\propto x$ 
denotes the hole concentration.  

$\kappa_{p}\simeq 0$ in the 
insulating phase.  The plane of the spin lattice in the lightly doped 
Mott insulator seems to be incompressible as indicated by the arrow 
for $x=6.0-6.38$.  The magnetic exchange energy $J$ of order 1500 K 
determines the electron-electron interactions in the Cu2 planes.  

$\kappa_{p}> 0$ in the metallic phases.  While weakly decreasing in 
the underdoped $\alpha$\/-ortho regime, $B(x)$ starts to collapse in 
the overdoped $\beta$\/-ortho regime.  The plateau between $x=6.8$ and 
6.9 points to phase segregation, possibly into incommensurate stripe 
phases.  In the metallic phase the doped holes hop with a $nn$ matrix 
element $t\propto d^{-n}$, $n\geq 2$, and thus shorter $nn$ distances, 
$d$, increase strongly the kinetic energy of the holes.  But the 
physics of the doped Mott insulator is that of competition between the 
exchange energy $J$ and the kinetic energy per hole $n_{h}t$.  The 
doped holes in the underdoped regime appear only as vacancies in the 
background of a spin singlet liquid.  The lattice of such a strongly 
correlated $t-J$ type electron system is expected to be much harder 
than that of a nearly noninteracting electron liquid.

Hence the $\alpha$\/- and $\beta$\/-ortho deformations may be 
identified as characteristic lattice responses to fundamentally 
different types of electron liquids in the metallic phase: a weakly 
compressible $t-J$\/ like in the under- and optimum doped, and a 
strongly compressible Fermi liquid-like in the overdoped regimes.

\subsection{Perpendicular O2,O3 displacements}

Figure~3 displays the doping dependence of the interlayer 
spacings Y--O2, Y--O3 from samples without \cite{CavRup} and with 
\cite{HewJil} a $c$\/-axis minimum.  At the onset of the metallic phase 
the degenerate tetragonal positions of the planar oxygens are split 
into O2 along $a$, and O3 along $b$. Both are also displaced along $c$\/: 
O2 by $\sim -0.05$ {\AA} below, and O3 by $\sim +0.05$ {\AA} 
above the tetragonal reference value, $cf.$ figure~4b.  
The anisotropic displacement of the planar oxygens along $c$ is usually attributed 
to the anisotropic Coulomb repulsion between the 1-dimensional charge 
reservoir and the 2-dimensional planes.

\begin{figure}[t]
\tabcapfont
\centerline{
\begin{tabular}{c@{\hspace{1.5cm}}c}
\includegraphics[width=5cm]{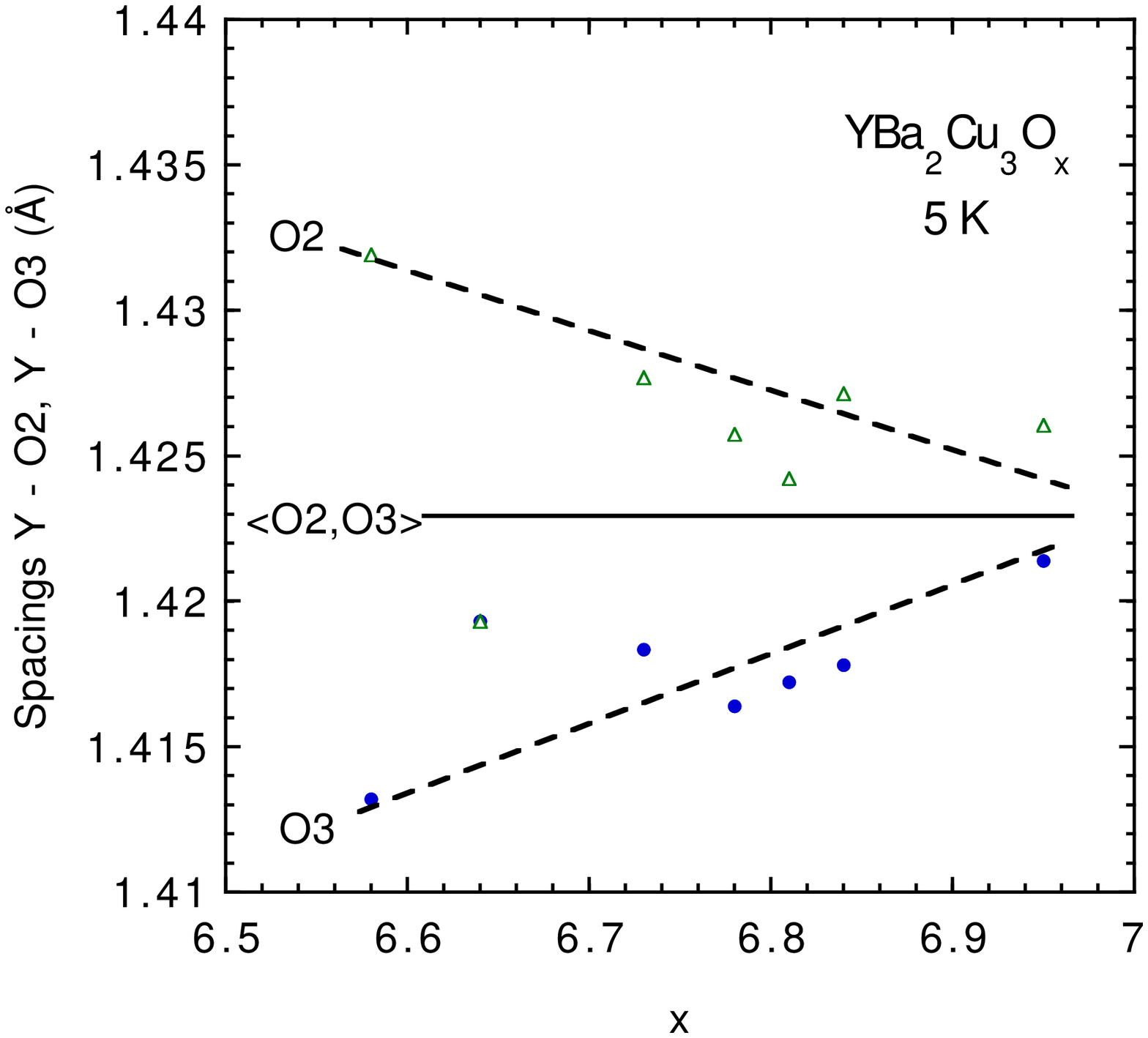} &
\includegraphics[width=5.cm]{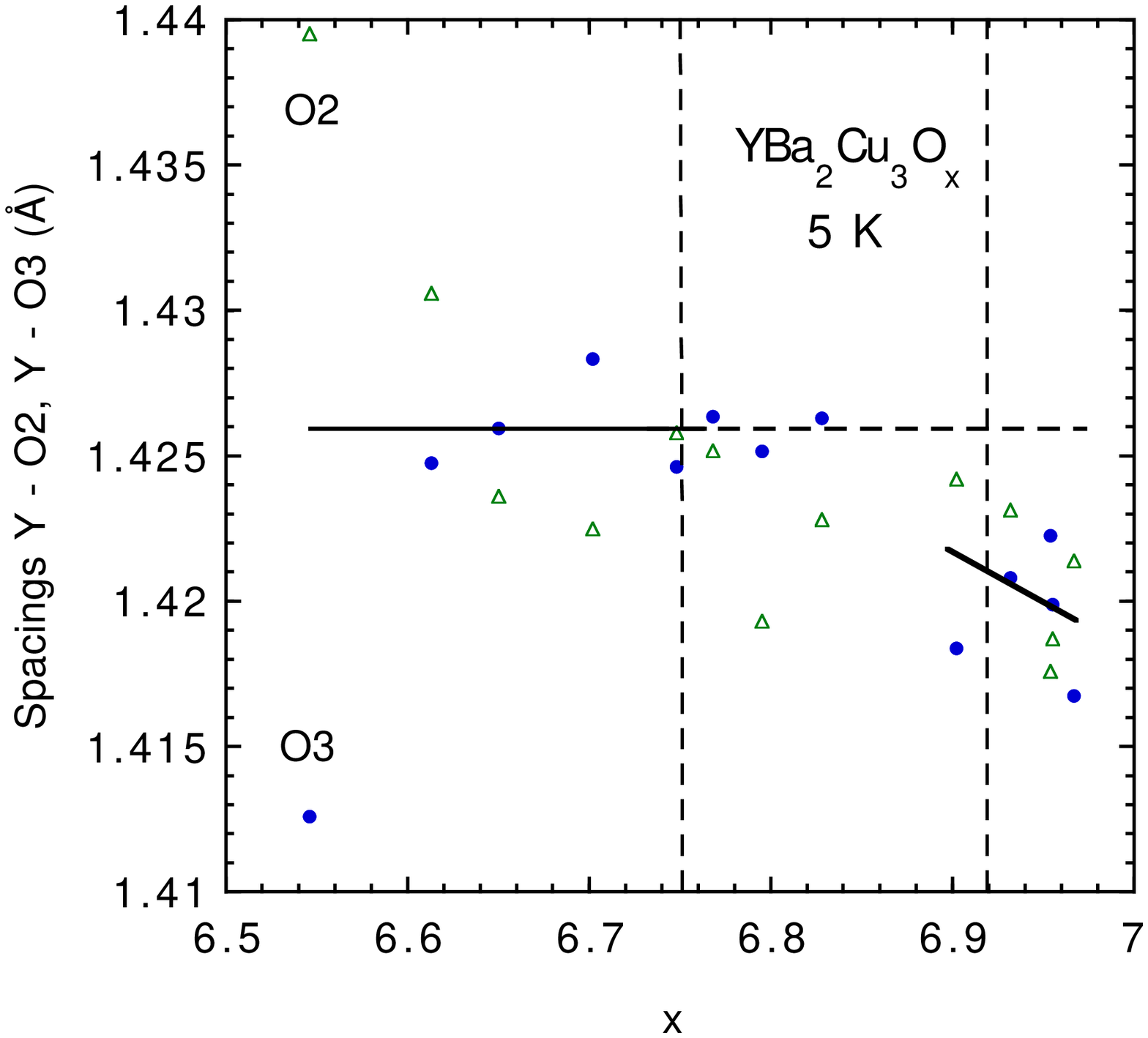} \\
a.~~ Without $c$\/-axis minimum (3) & b.~~ With $c$\/-axis minimum (13, 14)
\end {tabular}}
\caption{Spacings Y--O2, Y--O3 $vs.$ oxygen content in 
YBa$_2$Cu$_3$O$_x$ at 5 K from neutron diffraction in different 
samples. Triangles: O2 ($a$\/-axis). Dots: O3 ($b$\/-axis). All lines are guides to the eye. a.: 
In quenched samples from the carbonate route after Cava 
$et$ $al.$ (3). b.: In near equilibrium samples from the BaO route after 
Kaldis (13), Hewat $et$ $al.$ (14).} \label{oxygen}
\end{figure}
\begin{figure}[t]
\tabcapfont
\centerline{
\begin{tabular}{c@{\hspace{1.5cm}}c}
\includegraphics[width=5cm]{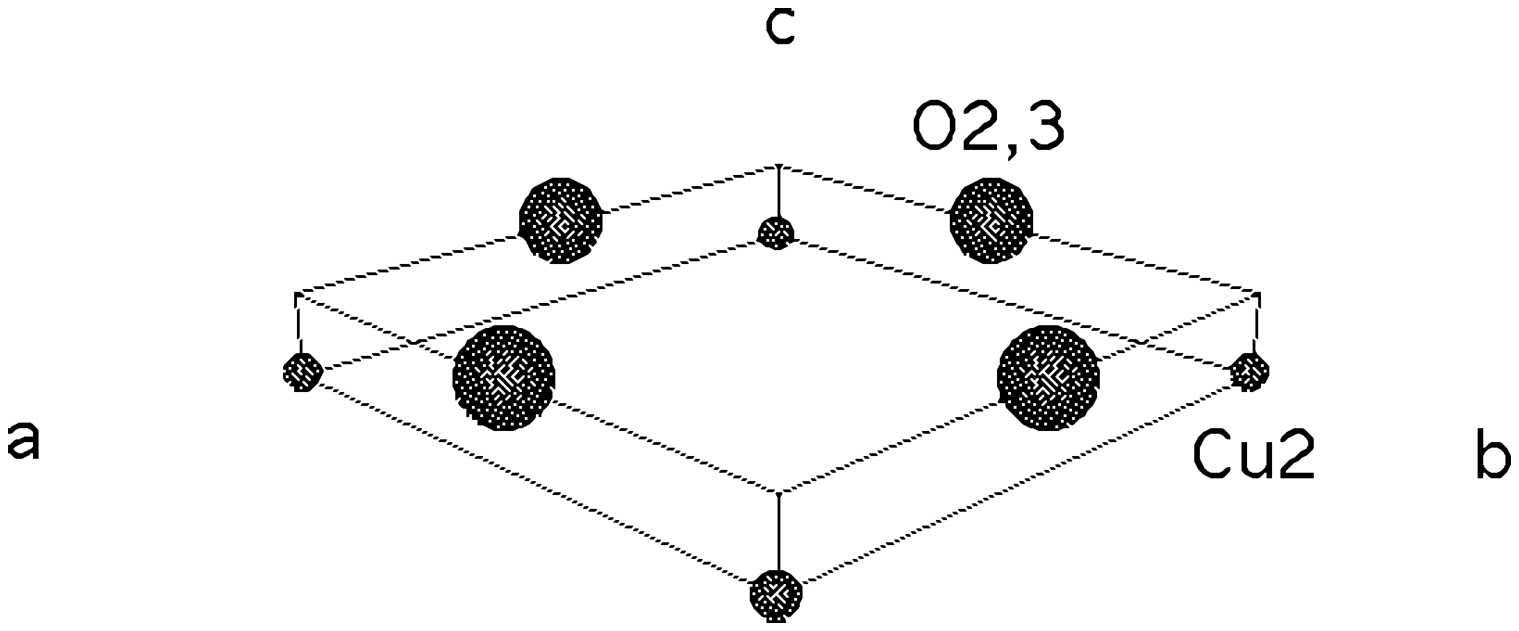} &
\includegraphics[width=5cm]{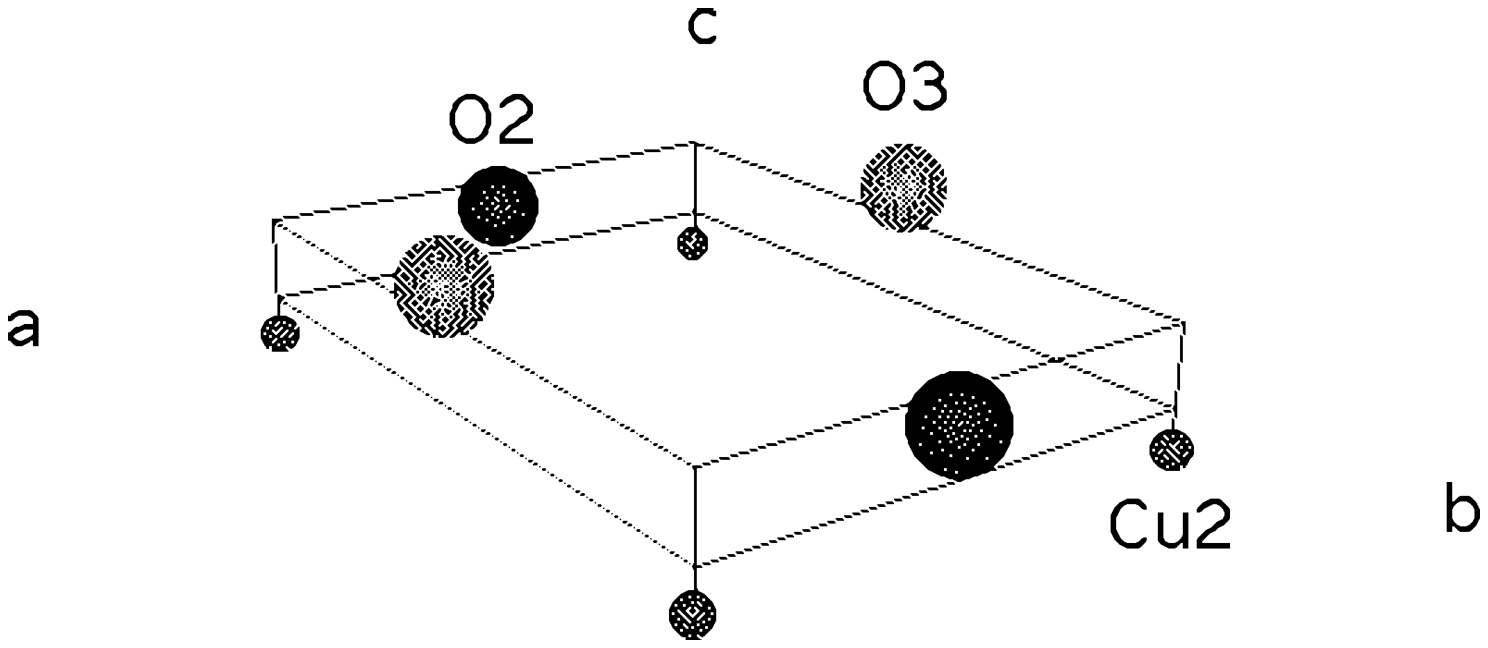} \\
a.~~ Tetragonal insulator  & b.~~ $\alpha$\/-ortho underdoped metal
\end {tabular}}
%
\bigskip
\centerline{
\begin{tabular}{c@{\hspace{2.3cm}}c}
\includegraphics[width=5cm]{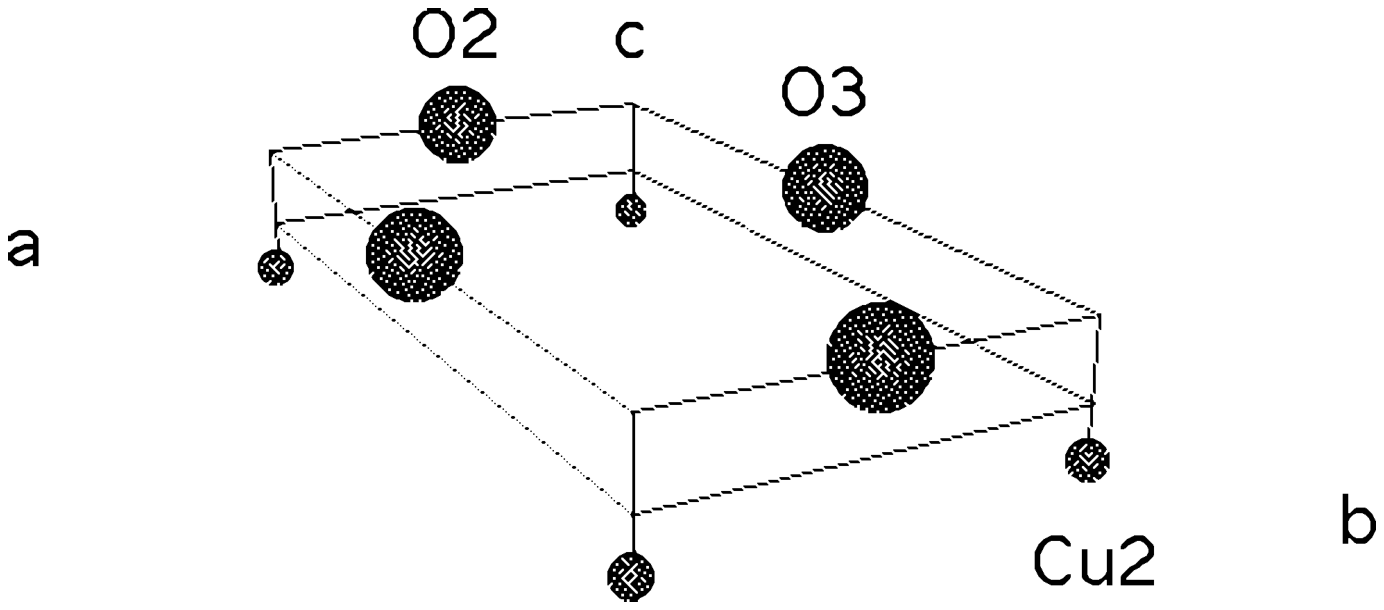} &
\includegraphics[width=4.3cm]{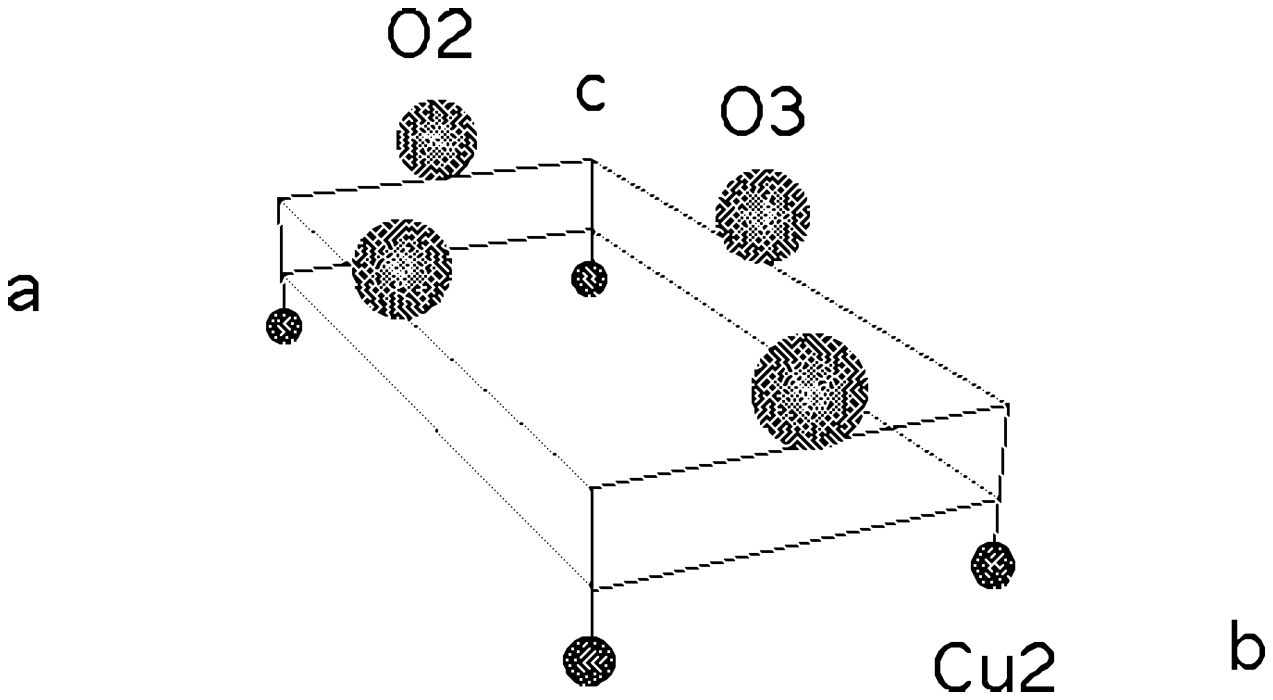} \\
c.~~ Optimum doped metal & d.~~ $\beta$\/-ortho overdoped metal
\end {tabular}}
\caption{Sketches of the CuO$_2$ lattices in YBa$_2$Cu$_3$O$_x$. 
The frame indicates the dimpling due to the surrounding chemistry and bandstructure 
effects.}
\label{planes}
\end{figure}
This is yet another example for Coulomb repulsion and hybridization 
are not able to correctly describe the doping-induced displacements in 
the CuO$_2$ layer: increasingly perfected chains in the reservoir 
layer relax the $c$\/-axis anisotropy of the planar O2, O3 instead 
of increasing them.  Thus the average spacing Y--$<$O2,O3$>$ matches the 
degenerate tetragonal reference value throughout the underdoped 
$\alpha$\/-ortho regime (thick drawn out line in figures~3a,b). 
Meaningful crystallographic studies of the overdoped $\beta$\/-ortho 
regime require samples with a $c$\/-axis minimum around $x_{opt}$ 
\cite{Kal01}.  Figure~3b displays the O2, O3 $c$\/-axis 
displacements in such samples \cite{Kal01,HewJil}.  Herein the transition 
into the overdoped regime seems to be connected with weak 
$unidirectional$ displacements of both oxygens O2, O3 towards the 
Y-layer, $cf.$ figure~4d.  Studies of the local 
atomic structure by Y-EXAFS \cite{RoeLoe,Kal01} confirm this unidirectional 
$c$\/-axis shift of the planar oxygens in the overdoped regime also in 
standard samples without $c$\/-axis minimum \cite{ConJil}.

We summarize the doping-induced displacements of 
Cu2 and O2,O3 in the schematic figures~4a-d:

$i.$ Metallic hole concentrations increase the 
two-dimensional density of Cu2 by dimpling the planes. There is some similarity 
with the mechanism collapsing an umbrella \cite{Roe00a}. As Cu2 moves 
out of the plane the basal area shrinks.

$ii.$ The underdoped regime is governed by the quadrupolar 
$\alpha$\/-ortho strain. O2 and O3 shift oppositely along $c$\/ 
such that the basal Cu2 area may adjust to a relative maximum.

$iii.$ Close to optimum doping the quadrupolar $\alpha$\/-ortho strain 
vanishes. O2 and O3 may achieve nearly degenerate $c$\/-axis positions.

$iv.$ In the overdoped regime the deformations have changed from the 
quadrupolar $\alpha$\/- to the monopolar $\beta$\/-ortho type 
stressing the $a$\/-, $b$\/-axes in the same direction therewith 
collapsing the basal Cu2 area. Both 
oxygens O2,3 shift perpendicularily in the $same$ direction along $c$\/.


%
\section{Concluding Remarks}

We have undertaken an attempt to identify among the many doping-driven 
lattice effects in YBa$_2$Cu$_3$O$_x$ those which are most likely 
connected with the generic low energy electronic structure.  Fermi 
surface driven lattice effects are well known from the classical 
metals.  For instance about 3/4 of the metallic elements tend to 
maximize the gain of kinetic energy by crystallization in most closely 
packed structures $hcp, fcc, bcc$.  In electron compounds and 
Hume-Rothery alloys the Fermi energy is well known to avoid maxima in 
the density of states and to drive structural transformations changing 
appropriately the symmetry of the Brillouin zone. These mechanisms 
are expected to be operative also in low dimensional and strongly 
correlated electron systems.  The competition between exchange and 
kinetic energies in strongly correlated electron systems however gives 
rise to new and more intricate lattice effects which, once disentagled 
from bandstructure effects, may yield new insights into the 
many electron ground state.

\begin{acknowledgements}
I thank E. Kaldis for a stimulating and fruitful cooperation. The ESRF 
supported this work partially through project HE731 at BM29.
\end{acknowledgements}

\end{article}

\end{document}